\documentclass[iop]{emulateapj}
\usepackage{color}
\usepackage{graphicx}
\usepackage{amsmath}
\usepackage{color}
\usepackage[section]{placeins}

\shorttitle{Testing the cosmic opacity at higher redshifts}
\shortauthors{Liu et al.}

\begin{document}
\title{Testing the cosmic opacity at higher redshifts: implication from quasars with
available UV and X-ray observations}

\author{Tonghua Liu\altaffilmark{1}, Shuo Cao\altaffilmark{1$\ast$}, Marek
Biesiada\altaffilmark{2$\dag$}, Yuting Liu\altaffilmark{1}, Shuaibo
Geng\altaffilmark{1}, and Yujie Lian\altaffilmark{1}}

\altaffiltext{1}{Department of Astronomy, Beijing Normal University,
Beijing 100875, China;\emph{caoshuo@bnu.edu.cn}}
\altaffiltext{2}{National Centre for Nuclear Research, Pasteura 7,
02-093 Warsaw, Poland; \emph{Marek.Biesiada@ncbj.gov.pl}}

\begin{abstract}

In this paper, we present a cosmological model-independent test for
the cosmic opacity at high redshifts ($z\sim5$). We achieve this
with the opacity-dependent luminosity distances derived from
nonlinear relation between X-ray and UV emissions of quasars,
combined with two types of opacity-independent luminosity distances
derived from the Hubble parameter measurements and simulated
gravitational wave (GW) events achievable with the Einstein
Telescope (ET). In the framework of two phenomenological
parameterizations adopted to describe cosmic opacity at high
redshifts, our main results show that a transparent universe is
supported by the current observational data at 2$\sigma$ confidence
level. However, the derived value of the cosmic opacity is slightly
sensitive to the parametrization of $\tau(z)$, which highlights the
importance of choosing a reliable parametrization to describe the
optical depth $\tau(z)$ in the early universe. Compared with the
previous works, the combination of the quasar data and the $H(z)$/GW
observations in similar redshift ranges provides a novel way to
confirm a transparent universe ($\epsilon=0$ at higher redshifts
$z\sim 5$), with an accuracy of $\Delta \epsilon\sim 10^{-2}$. More
importantly, our findings indicate that a strong degeneracy between
the cosmic opacity parameter and the parameters characterizing the
$L_{UV}-L_X$ relation of quasars, which reinforces the necessity of
proper calibration for such new type of high-redshift standard
candle (in a cosmological model-independent way).

\end{abstract}


\maketitle


\section{Introduction}

It is well recognized in modern cosmology that our universe is
undergoing an accelerated expansion at the current stage, and this
conclusion is supported by the most direct evidence that type Ia
supernovae (SNe Ia) are unexpected observed to be fainter than
expected in a decelerating universe
\citep{Riess98,Perlmutter99,Scolnic18}. Besides a new cosmological
component exerting negative pressure (cosmological constant being
the simplest candidate)
\citep{Ratra88,Caldwell98,Cao11a,Cao13a,Cao14,Ma17,Qi18}, some other
different theories or mechanisms have also been proposed to explain
the observed SNe Ia dimming \citep{Qi17,Xu18}. Indeed, soon after
the discovery of cosmic acceleration, it was suggested that SNe Ia
observations could be affected by the non-conservation of photon
number in the beam of emitted light. From theoretical point of view
such so-called cosmic opacity may be due to many possible
mechanisms. The standard mechanism focuses on the photon absorbtion
or scattering by dust particles in the Milky Way, intervening
galaxies and the host galaxy \citep{Tolman30}. Some other exotic
mechanisms for cosmic opacity discuss the conversion of photons into
gravitons \citep{Chen95}, light axions in the presence of
extragalactic magnetic fields
\citep{Csaki02,Avgoustidis10,Jaeckel10}, or Kaluza-Klein modes
associated with extra-dimensions \citep{Deffayet00}. Up to now, more
than 1000 SNe Ia have been detected \citep{Scolnic18}, hence the
constraints on cosmological parameters inferred from SNe Ia are
currently more suffering from the systematic errors \citep{Qi19b}.
Cosmic opacity might be an important source of systematic errors in
this respect and it becomes increasingly important to quantify the
transparency of the Universe.

In the previous works, two general methods have been widely applied
to obtain model-independent constraints on the cosmic opacity,
through the combination of opacity-dependent luminosity distances
(LD) derived from SNe Ia observations and other sources of
opacity-independent cosmological distance observations. The first
method generally focus on angular diameter distances (ADD) inferred
from galaxy clusters or baryon acoustic oscillations (BAO)
\citep{More09,Chen12,Li13}, while the second approach is based on
the opacity-independent luminosity distances reconstructed by the
Hubble parameter ($H(z)$) measurements of differential ages of
passively evolving galaxies --- the so called cosmic chronometers
\citep{Holanda13,Liao13,Liao15,Jesus17,Wang17}. It is worth noting
that the first method relies on a fundamental relation known as
``distance duality relation" (DDR)
\citep{Etherington1,Etherington2,Cao11}, which states that LD and
ADD are quantitatively related as
$D_\mathrm{L}=D_\mathrm{A}(1+z)^2$, where $D_\mathrm{L}$ and
$D_\mathrm{A}$ are respectively the LD and ADD at the same redshift
$z$. This relation is valid in any metric theory of gravity,
provided the number of photons is conserved within the beam
\citep{Ellis}. As was pointed out by \citep{Avgoustidis10}, an
opacity parameter $\tau$ can be introduced to describe the optical
depth associated with cosmic absorption. In the case of standard
candles, the flux received by the observer will be reduced by a
factor of $e^{-\tau(z)}$ and the observed LD
($D_{\mathrm{L,obs}}(z)$) should be rewritten as
\begin{equation}{\label{eq3}}
D_{\mathrm{L,obs}}(z)  = D_{\mathrm{L,true}}(z) \cdot e^{\tau(z)/2}.\;
\end{equation}
where $D_{\mathrm{L,true}}(z)$ represents the true luminosity
distance. The photon number in the beam from a source at $z$ to an
observer at $z=0$ will be reduced when $\tau(z)>0$.

Since the nature of cosmic opacity is unknown, it is still an open
question whether it can be partly responsible for the dimming of
standard candles at higher redshifts. In this work, in order to test
the transparency of the universe spanning a wide redshift range, we
will use the quasar data compiled by \cite{Risaliti2018} as the
source of LDs potentially affected by the opacity ($z\sim5$). These
luminosity distances were proposed to be assessed from the nonlinear
relation between the ultraviolet (UV) and X-ray luminosity
\citep{Risaliti2015,Zheng20,Liu20}. On the other hand, it is also
rewarding to have objects with measurable luminosity distances
spanning a wide redshift range. Therefore, two types of opacity-free
distance probes will also be included in our analysis, i.e., 31
Hubble parameter measurements from differential ages of passively
evolving galaxies
\citep{Jimenez03,Simon05,Stern10,Moresco12,Zhang14,Moresco15,Moresco16}
and simulated gravitational waves (GWs) based on the future
generation of GW detector -- the Einstein Telescope (ET)
\citep{Qi19b,Wei19,Dalal2006,Taylor12,Zhao11,Cai16}. Specially, the
waveform signals of GWs from inspiralling and merging compact
binaries have the advantage of being insensitive to the
non-conservation of photon number \citep{Abbott16,Abbott17}.
Therefore, GW standard sirens provide a new way to determine the
opacity-independent luminosity distances of quasars at the same
redshifts \citep{Schutz86}, because GWs propagate freely through a
perfect fluid without any absorption and dissipation, unaffected by
the opacity of the universe. Note that in the framework of the
third-generation ground based GW detectors, the observational
constraints on the cosmic opacity can be extended to the distant
universe ($z\sim5$), with a large amount of high-redshift inspiral
events registered per year at its design sensitivity \citep{ET}.
Hence, the so-called ``redshift desert'' problem could be alleviated
to some extent with respect to the exploration of the cosmic
opacity.

The paper is organized as follows. In Section 2, we present the
observational data used in our work, including the quasars sample,
Hubble parameter measurements and the simulated GW sample. In
Section 3, we introduce the phenomenological parameterizations of
the cosmic opacity $\tau(z)$, and furthermore present the unbiased
cosmic-opacity constraints. Finally, the main conclusions are
summarized and discussed in Section 4.

\section{observational and simulated data}

In this section, we will briefly introduce the methodology of
deriving two different types of cosmological distances, i.e.,
opacity-dependent luminosity distances from quasar measurements in
X-ray and UV bands (standard candles), as well as
opacity-independent luminosity distance inferred from the current
$H(z)$ data (cosmic chronometers) and future GW observations
(standard sirens).

\subsection{Opacity-dependent luminosity distances from quasars}

In order to measure the luminosity distance at cosmological scales,
one always turns to the most luminous sources of known (or
standardizable) intrinsic luminosity. Quasars have always been
considered as very promising candidates for this role, since they
can be observed up to very high redshifts and are one of the
brightest sources in the Universe. Considering the extreme
variability and a wide luminosity range in quasars, many previous
studies have used different relations to indirectly derive their
luminosity distances, i.e., the Baldwin effect \citep{Baldwin77},
the Broad Line Region radius - luminosity relation \citep{Watson11},
and the properties of highly accreting quasars \citep{Wang13}. In
this paper, we make use of the recent progress made in the nonlinear
relation between the UV and X-ray luminosities of quasars
\citep{Avni1986}, which provides an effective approach to construct
the Hubble diagram beyond the redshift limit of SNe Ia
\citep{Risaliti2015}.

In the framework of a scenario generally recognized in the studies
of quasars, the UV photons are emitted by an accretion disk and the
X-rays are generated by Compton upscattered photons from an
overlying hot corona. Therefore, the non-linear relation between
luminosities in the X-rays ($L_X$) and UV band ($L_{UV}$) can be
expressed as
\begin{equation} \label{X-UV relation}
\log(L_X)=\gamma\log(L_{UV})+\beta',
\end{equation}
where $L_X$ and $L_{UV}$ represent the rest-frame monochromatic
luminosity at 2keV and 2500${\AA}$ ($\log=\log_{10}$), while
$\beta'$ and $\gamma$ denote the intercept and the slope parameter,
respectively. Such relation can be rewritten in terms of the fluxes
$F$, the slope $\gamma$ and the normalization constant $\beta$:
\begin{equation} \label{QSO LD}
\log(D_L)=\frac{1}{2(1-\gamma )}\times[\gamma\log(F_{UV})-\log(F_X)
+ \beta ],
\end{equation}
where $\beta$ is a constant that quantifies the slope and the
intercept, $\beta=\beta'+(\gamma-1)\log 4\pi$.

Unfortunately, it should be pointed that quasars always display a
very high intrinsic scatter in the observed relation, which remains
one of the major disadvantages in measuring quasar distances with
high precision in this approach. Based on the parent sample
of 1138 quasars from the COSMOS survey \citep{Lusso2010} and the
Sloan Digital Sky Survey (SDSS) and XMM spectra \citep{Young2010},
\citet{Risaliti2015} built the ``best sample" of 808 quasars
suitable for cosmological applications, focusing on the difference
between the intrinsic and observational dispersion in the
$L_{UV}-L_X$ relation. More specifically, it was found that when
quality filters are applied to the parent quasar sample (discarding
the broad absorption line and radio-loud quasars), the dispersion
previously estimated at the level of $\delta=0.35-0.40$ will be
significantly reduced, with the derived value of the slope parameter
$\gamma=0.6\pm0.02$ and the dispersion $\delta=0.3$. The subsequent
studies indicated that the observed dispersion could quantify two
distinct contributions: a scatter due to various observational
effects and an intrinsic scatter related to unknown physics
\citep{Lusso2016,Risaliti2017,Bisogni17,Risaliti2018}. Following
this direction, a sub-sample of quasars was identified without the
major contributions from uncertainties in the measurement of the
(2keV) X-ray flux, absorption in the spectrum in the UV and in the
X-ray wavelength ranges, variability of the source and
non-simultaneity of the observation in the UV and X-ray bands,
inclination effects affecting the intrinsic emission of the
accretion disc, and the selection effects due to the Eddington bias
\citep{Risaliti2017}. Great efforts have also been made to quantify
the observational effects, with the intrinsic dispersion attested to
be smaller than 0.15 dex \citep{Risaliti2018}.

Besides the dispersion in the $L_{UV} - L_X$ relation, the
reliability and effectiveness of the method also strongly depend on
its possible evolution with redshifts, which should be checked on
quasar sub-samples in different redshift intervals
\citep{Bisogni17}. Taking the above effects into full consideration,
\citet{Risaliti2018} collected a sample of 1598 high-quality quasars
with available X-ray and UV measurements, based on the
cross-correlation of the XMM-Newton Serendipitous Source Catalogue
Data Release 7 \citep{Rosen2016} and the Sloan Digital Sky Survey
(SDSS) quasar catalogues from Data Release 7 \citep{Shen11} and Data
Release 12 \citep{Paris17}. With the gradually refined selection
technique and flux measurements, as well as the elimination of
systematic errors caused by various aspects, the final results
showed that such quasar sample could produce a Hubble diagram in
excellent agreement with the concordance $\Lambda$CDM model and the
SNe Ia sample in the redshift range of $z<1.4$, with accurate
determination of the slope parameter $\gamma=0.6\pm0.02$ and smaller
dispersion $\delta=0.24$. Therefore, new measurements of the
expansion rate of the universe could be performed in the range
$0.036<z<5.100$, which has never been explored before by any other
cosmological probe. The intercept parameter $\beta$ still needs to
be calibrated by using an external calibrator like SNe Ia. Such
analysis has already been implemented in \citet{Risaliti2015}, which
seek constraints on the intercept $\beta$ and cosmological
parameters in the non-flat $\Lambda$CDM model, using a smaller
sample of quasars data available. It is worth noting that even
$\beta$ is considered as a free parameter in the cosmological fit,
the cross-calibration is still necessary since both SNe and quasars
are used as ``standardized candles" and the cosmological parameters
are obtained from the same physical quantity, i.e., the distance
modulus. Following the approach analogous to what is done
with Type Ia SNe, \citet{Melia19} used the recently assembled quasar
sample \citep{Risaliti2018} to compare the predictions of two
distinct cosmological models ($R_h=ct$ and $\Lambda$CDM
cosmologies). Along with the free parameters of the cosmology
($\Omega_m$) itself, parameters characterizing the nonlinear
relation between X-ray and UV emissions of quasars ($\gamma$,
$\beta$, and $\delta$) are simultaneously constrained in a global
fitting on the quasar data alone. Such methodology, which generated
the best-fit parameters of $\Omega_m=0.31\pm 0.05$, $\gamma=0.639\pm
0.005$, $\beta=7.02\pm 0.012$ and $\delta=0.231\pm0.0004$ in the
$\Lambda$CDM model, still requires the use of model-dependent
luminosity distances to provide the inferred luminosities and thus
the constants quantifying the $L_{UV}-L_X$ relation of quasars.

Finally, it should be emphasized that the cosmic opacity could
affect the LD measurements of quasars, as displayed in Eq.~(1). From
this point of view, one might gain hints whether the dispersion of
the flux measurements are affected by the X-ray/UV absorption
unaccounted for in the data reduction process (even though dust
extinction has been carefully considered in the sample selection).

\subsection{Opacity-independent luminosity distances from cosmic chronometers}

For the current observations of Hubble parameter, we turn to the
newest compilation covering the redshift range of $0.01<z<2.3$
\citep{Jimenez03,Simon05,Stern10,Moresco12,Zhang14,Moresco15,Moresco16},
which consists of 31 cosmic chronometer $H(z)$ data and 10 BAO
$H(z)$ data (see \citet{Zheng16,Wu20} for more details). However,
one should be aware that the BAO method is
cosmological-model-dependent, i.e., the measurements based on the
identification of BAO and the Alcock-Paczy\'{n}ski distortion from
galaxy clustering depend on how ``standard rulers" evolve with
redshift \citep{Blake12}. On the other hand, the cosmic chronometers
approach resulting in $H(z)$ measurements is
cosmological-model-independent, since the observations of cosmic
chronometers via differential ages of passively evolving galaxies
are independent of any specific cosmological model. Therefore, in
our work we consider only the cosmic chronometer $H(z)$ data to
reconstruct the $H(z)$ function and then derive opacity-independent
luminosity distances ($D_L(z)$) within the redshift range of
$0<z<2.0$. Specially, in the framework of the FLRW metric in a flat
universe, the luminosity distance can be written as
\begin{equation}\label{HzLD}
D_{L,CC}(z)=c \; (1+z)\int_0^z\frac{dz'}{H(z')},
\end{equation}
where $c$ is the speed of light and $H(z)$ is the Hubble parameter
at redshift $z$. Here, we use a simple trapezoidal rule to calculate
$D_{L}$ function, the uncertainty associated to the $i_{th}$
redshift bin is given by
\begin{equation}
s_i={\frac{c}{2}}(z_{i+1}-z_i)\left({\sigma_{H_{i+1}}^2\over
H_{i+1}^4} + {\sigma_{H_{{i}}}^2\over H_{i}^4}\right)^{1/2}\;,
\end{equation}
where $\sigma_{H_{i}}$ is the uncertainty of the $H(z)$ data.
Following the recent works of
\citet{Cao17a,Cao17b,Cao18,Cao19,Qi19a,Liu19,Zheng19}, a
nonparametric smoothing method of Gaussian Processes (GP)
\citep{Seikel12a} will be applied to achieve the above
reconstructions with the opacity-free expansion rate measurements.
Our analysis is based on the publicly available code called the GaPP
(Gaussian Processes in Python)
\footnote{http://www.acgc.uct.ac.za/~seikel/GAPP/index.html}. We
refer the reader to \citet{Holsclaw10,Qi18,Wu20} for more details
about GP reconstructed distances with $H(z)$ data.

Moreover, two recent measurements of the Hubble constant:
$H_0=67.4\pm0.5$ km/s/Mpc with 0.7\% uncertainty \citep{Planck18}
and $H_0=73.24\pm1.74$ km/s/Mpc with 2.4\% uncertainty (local $H_0$
measurement) \citep{Riess16} will be adopted in the distance
reconstruction with the GP method. The reconstructed $H(z)$ and
$D_{L}(z)$ functions and the corresponding 1$\sigma$ uncertainty
strips are shown in Fig.~1.

\begin{figure}[tbp]

\includegraphics[height=1.6in, width=1.8in]{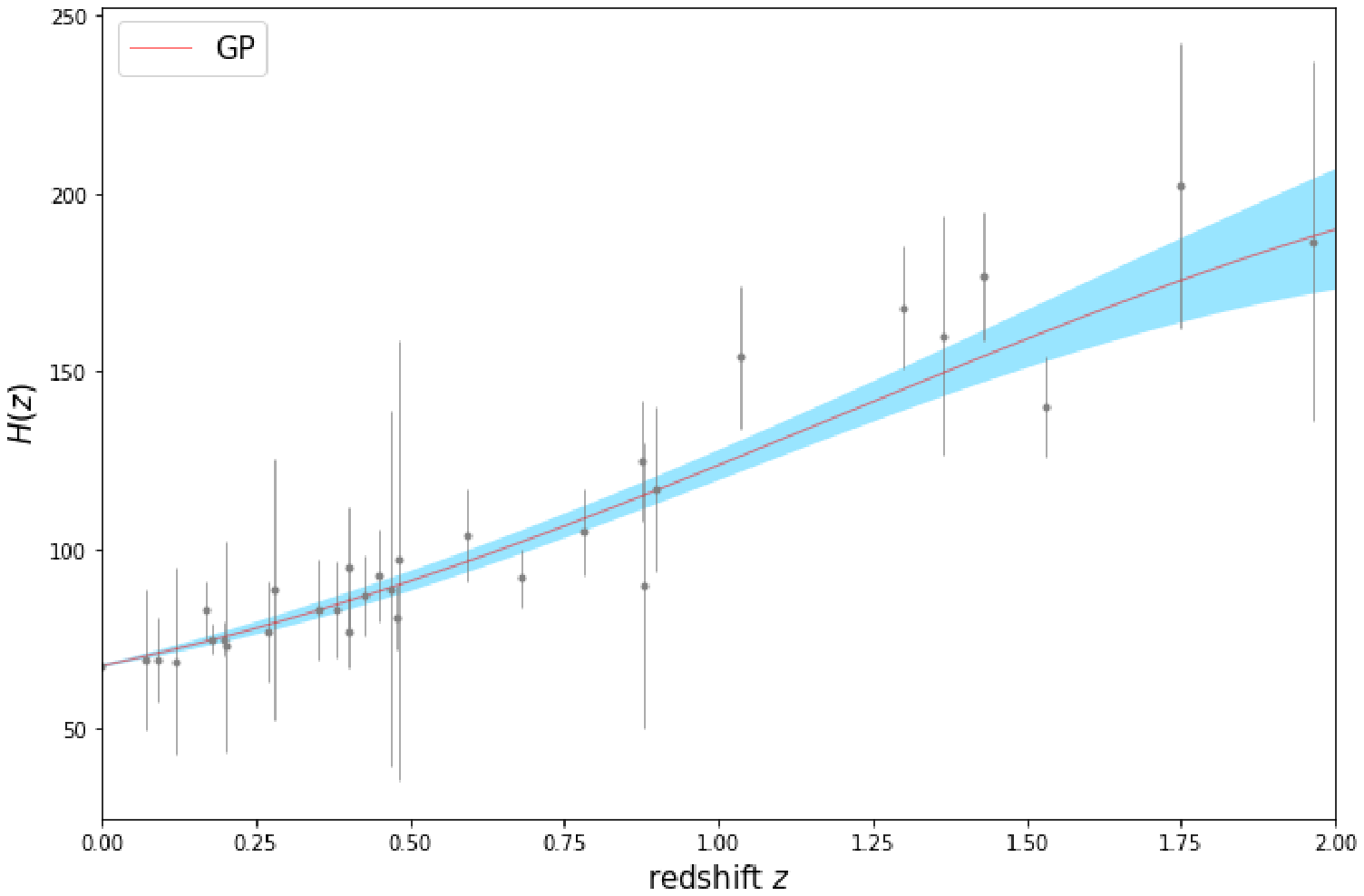}\includegraphics[height=1.6in, width=1.8in]{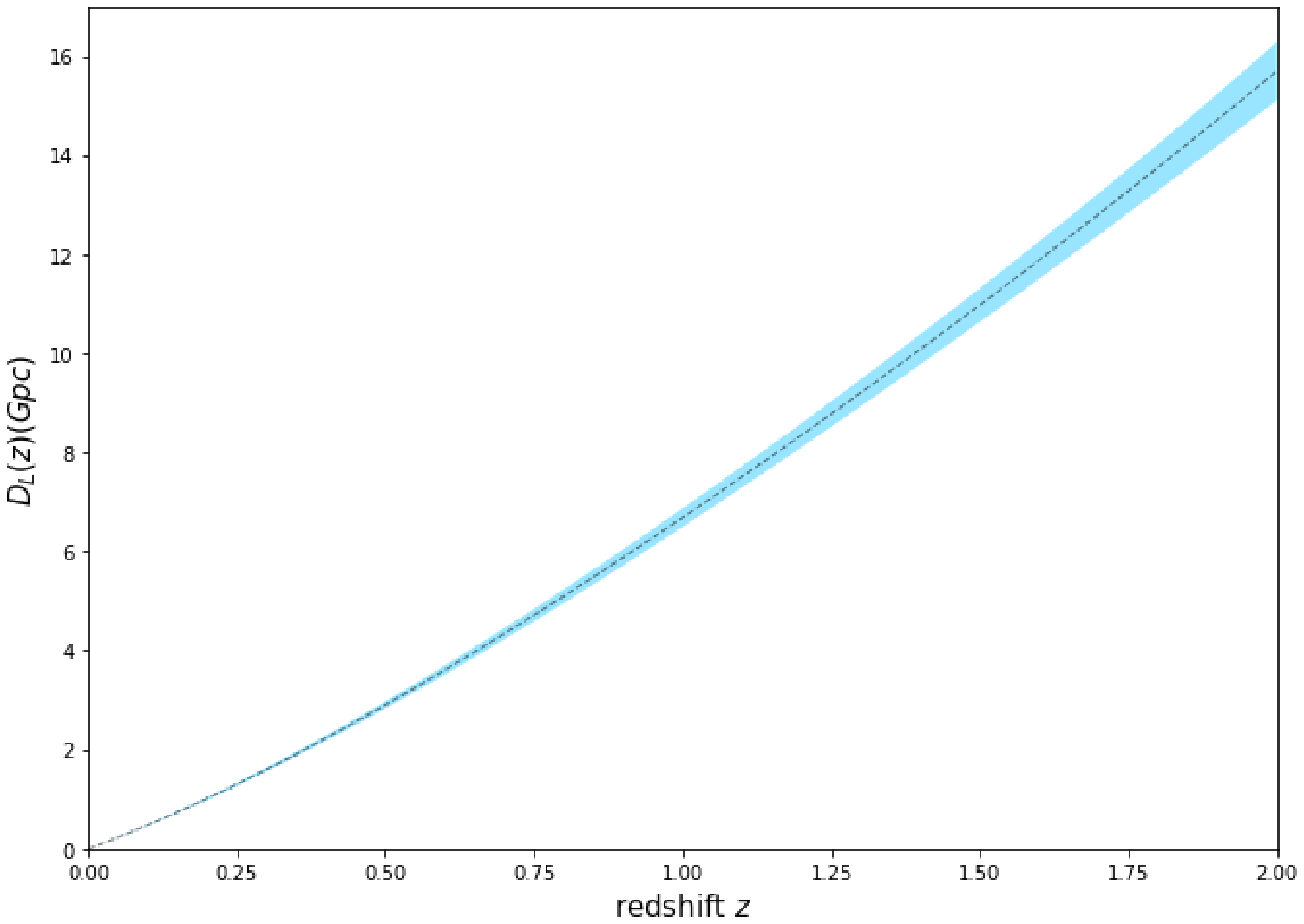}
\includegraphics[height=1.6in, width=1.8in]{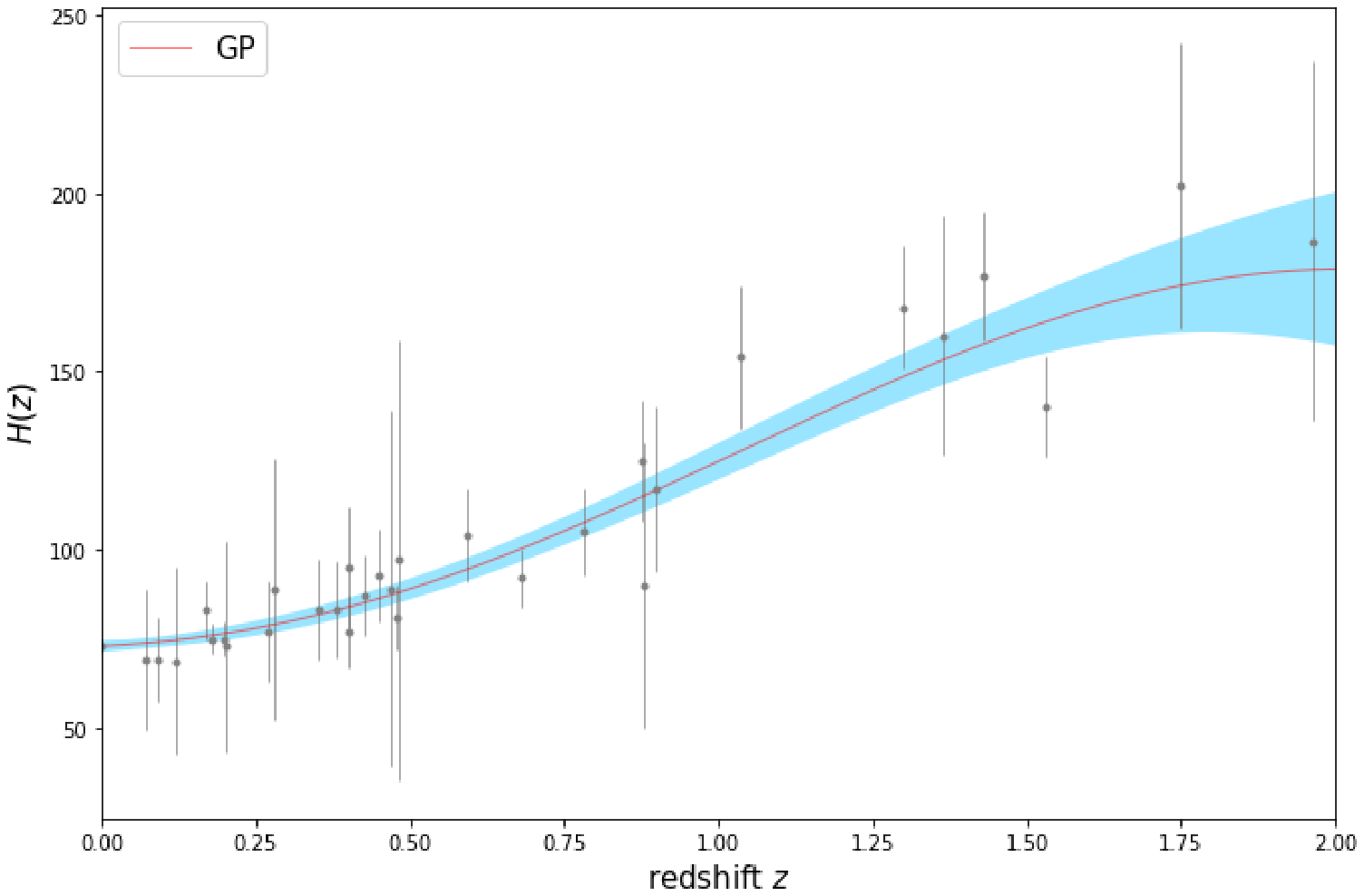}\includegraphics[height=1.6in, width=1.8in]{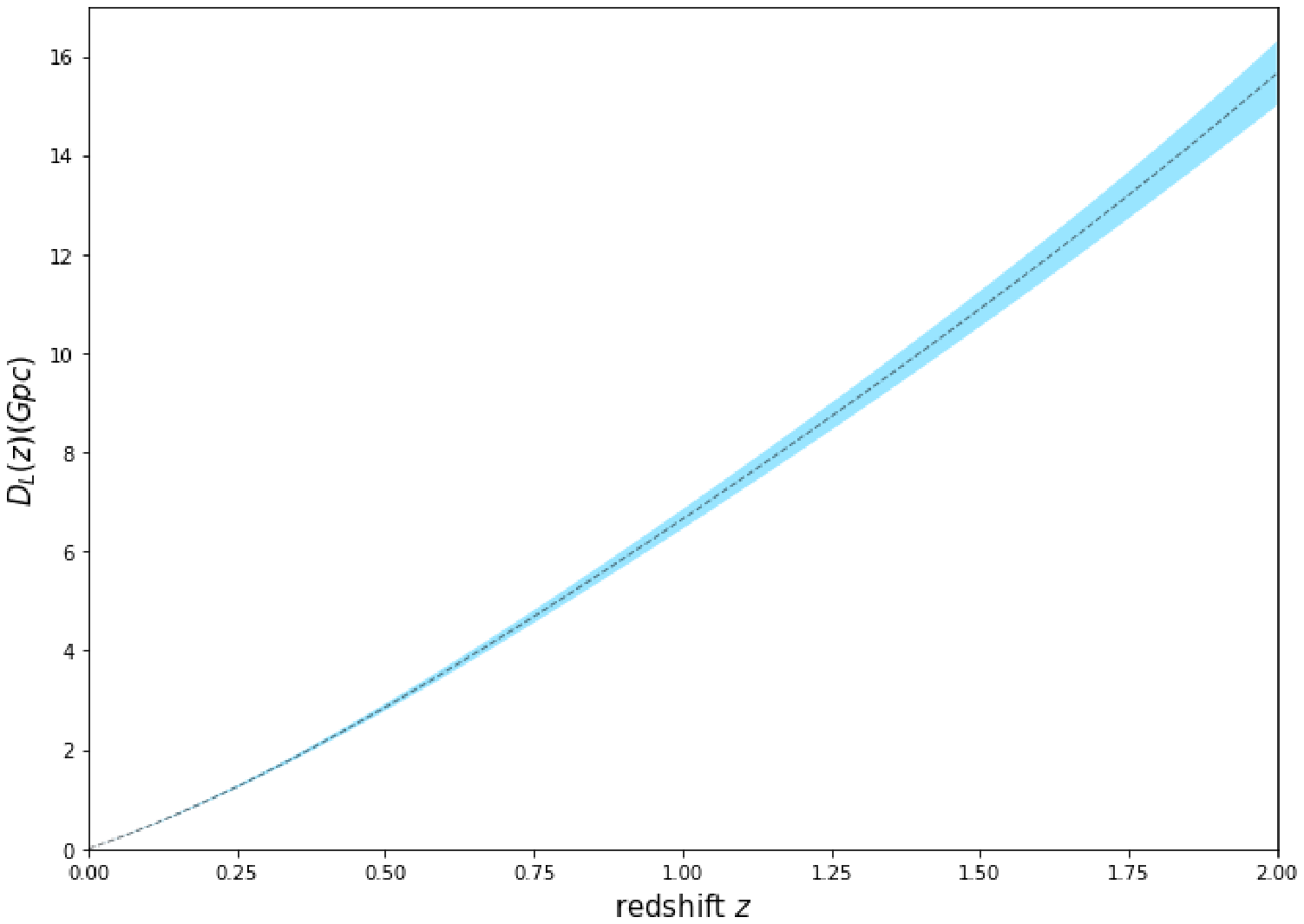}
\caption{The Gaussian Processes reconstruction of $H(z)$ function
(left panel) and $D_L(z)$ function (right panel). Two different
Hubble constant priors are adopted for comparison: $H_0=67.4\pm0.5$
km/s/Mpc (upper panel) and $H_0=73.24\pm1.74$ km/s/Mpc (lower
panel).} \label{HzGP}
\end{figure}

\begin{figure}
\begin{center}
\includegraphics[width=0.95\linewidth]{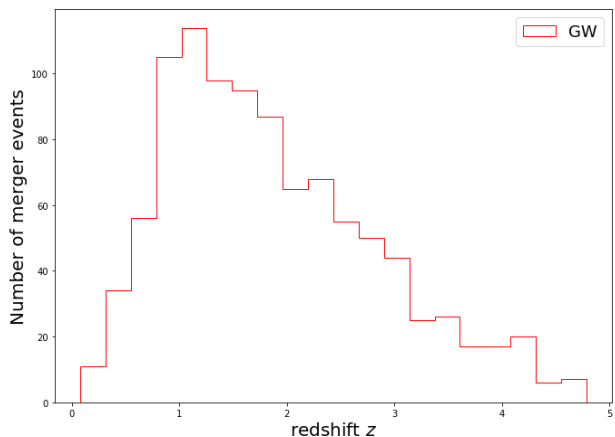}
\end{center}
\caption{The redshift distribution of 1000 simulated GW events from
ET.}
\end{figure}

\begin{figure}
\begin{center}
\includegraphics[width=0.95\linewidth]{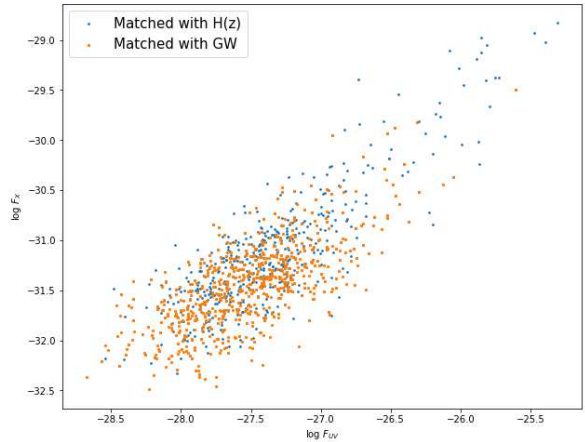}
\end{center}
\caption{The scatter plot of the UV and X-ray flux measurements in
the quasar sample used in the cosmic-opacity test.} \label{flux2}
\end{figure}

\subsection{Opacity-independent luminosity distances derived from simulated GWs}

Another opacity-independent luminosity distance can be derived from
the standard sirens, i.e., the gravitational wave signals from
inspiraling binary systems consisting of neutron stars (NSs) or
mixed neutron stars - black holes (NS-BHs) \citep{Schutz86}.
Moreover, if the source redshifts can be independently determined
from EM counterparts associated with the GW events, these standard
sirens will provide a model-independent way to obtain $D_L(z)$ over
a wide redshift range, well overlapped with that covered by the
quasars. Since the GWs travel in the universe without any absorption
or scattering by dust, their luminosity distance will not be
affected by cosmic opacity. We simulate GW signals based on the
foreseen performance of the third-generation gravitational wave
detector -- the Einstein Telescope \citep{ET}. Such ground-based
detector, which is designed to have a fantastic sensitivity in the
frequency range of $1-10^4$ Hz, will detect tens or hundreds of
thousand NS-NS inspiral events per year up to the redshift $z \sim
2$ and NS-BH mergers up to $z \sim 5$. Compared with the
advanced detectors such as AdLIGO and AdVirgo, such proposed
third-generation detector aims for a broadband factor of 10
sensitivity improvement, especially for the characteristic distance
sensitivity \citep{Taylor12b}. See \citet{Abernathy11} for the
details of the ET conceptual design study.

Inspired by this, we perform a Monte Carlo simulation of LDs
obtainable from the GW signals from NS-NS and NS-BH systems,
with sufficiently high signal to noise ratio (SNR) based on
the third generation technology (the advanced ``xylophone"
configuration) \footnote{For the characteristics of ET
considered in our analysis, the detector's noise curve is
approximated by the so called ``xylophone" configuration, with the
corresponding characteristic distance parameter $r_0=1918$ Mpc
\citep{Taylor12b}.}. In our simulation, we take the
transparent universe as the baseline model, in the framework of the
fiducial cosmological model (flat $\Lambda$CDM). Now the GW
luminosity distance can be written as
\begin{equation}
D_{L,GW}(z)=\frac{c(1+z)}{H_0}\int^z_0\frac{dz'}{\sqrt{\Omega_m(1+z')^3+(1-\Omega_m)}},
\end{equation}
with the matter density parameter and the Hubble constant fixed at
$\Omega_m=0.315$ and  $H_0=67.4$ km/s/Mpc from the latest Planck CMB
observations \citep{Planck18}.

Two sources of uncertainties are included in our simulation of the
luminosity distances from GW observations. Following the strategy
proposed in \citet{Cai16,Cai17}, the distance precision per GW event
is
\begin{eqnarray}
\sigma_{D_{L,GW}}=\sqrt{(\sigma_{D_{L,GW}}^{\rm
inst})^2+(\sigma_{D_{L,GW}}^{\rm lens})^2}, \label{sigmadl}
\end{eqnarray}
where the instrumental uncertainty is estimated as
$\sigma^{inst}_{D_{L,GW}}\simeq\frac{2D_{L,GW}}{\rho}$ and ${\rho}$
denotes the combined signal-noise ratio(SNR) for the three
independent detectors (a GW event is usually claimed when the
signal-to-noise ratio (SNR) of the detector network reaches above 8)
\citep{Zhao11}. Moreover, the lensing uncertainty due to the
line-of-sight (LOS) mass distribution is modeled as
$\sigma^{lens}_{D_{L,GW}}/D_{L,GW}=0.05z$ \citep{Li2015}. Following
the redshift distribution of GW sources taken after
\citet{Sathyaprakash2010} and sampling the the mass of neutron star
and black hole within [1,2] $M_{\odot}$ and [3,10] $M_{\odot}$, we
simulate 1000 GW events observable in the ET and their redshift
distribution is shown in Fig.~2. The specific method of simulating
the mock data of standard siren events is similar with
\citet{Qi19a,Qi19b}.

\section{Observational constraints on the cosmic opacity}

It should be recalled that, using X-ray and UV emission of quasars
to derive the luminosity distance (according to the $L_X - L_{UV}$
relation in Eq.~(\ref{QSO LD})) requires the knowledge of $\gamma$
and $\beta$ parameters. In this paper, we follow a common procedure
extensively applied in the SNe Ia cosmology, i.e., the light curves
of these standard candles are characterized by several ``nuisance''
parameters are optimized along with the unknown parameters of the
cosmological model \citep{Wei19,Liao15,Qi19b}. Similarly, in our
analysis the $\gamma$ and $\beta$ parameters will be fitted jointly
with the parameters associated with the cosmic opacity. In addition,
in order to avoid the corresponding bias of redshift differences
between the opacity-independent and opacity-dependent LDs, a
selection criterion ($|z_{QSO}-z_{CC/GW}|<0.005$) for a given pair
of data is used. The scatter plot of the quasar sub-sample are shown
in Fig.~3, which shares the same redshifts of the
opacity-independent luminosity distances from cosmic chronometers
and simulated GW events.

In order to constrain the cosmic-opacity parameter ($\tau$), we
perform Markov Chain Monte Carlo (MCMC) simulation to minimize the
$\chi^2$ objective function defined as
\begin{equation}
\chi^2 = \sum_{i=1}^{N} \frac{\left[\log(F_{X,i}) - \Psi_{th}
([F_{UV,i}];D_L[z_i],\tau(z_i),
\gamma,\beta)\right]^2}{\sigma_{F_{X,i}}^2+\sigma_{\Psi_{th,i}}^2+\delta^2},
\end{equation}
where $\sigma_{F_{X,i}}$ is the uncertainty of the $i$-th X-ray flux
measurement \footnote{Since the uncertainty of the UV flux
measurement $\sigma_{F_{UV,i}}$ is much lower than
$\sigma_{F_{X,i}}$ and $\delta$, we ignore it in the uncertainty
budget \citep{Risaliti2018}.}, and $\delta$ denotes the global
intrinsic dispersion (and we treat it as a nuisance parameter as
well). The function $\Psi_{th}$ that quantifies the theoretical
prediction of the X-ray flux from the UV flux is defined as
\begin{eqnarray}\nonumber
\Psi_{th}&=& \gamma \log(F_{UV,i})+2(\gamma-1)\log D_L(z_i)\\
&+&(\gamma-1)(\log e)\tau(z_i) + \beta,
\end{eqnarray}
where $D_L(z_i)$ is the LD calculated from the $ith$ reconstructed
Hubble parameter or the simulated GW data. The uncertainty
$\sigma_{\Psi_{th,i}}$ can be calculated through the error
propagation formula
\begin{eqnarray}
\sigma_{\Psi_{th,i}}=2(\gamma-1)\frac{\sigma_{D_{L,CC/GW}}}{\ln(10)D_{L,CC/GW}(z_i)},
\end{eqnarray}
where $\sigma_{D_{L,CC}}$ and $\sigma_{D_{L,GW}}$ are respectively
given by Eq.~(5) and (7). The summarization is performed over the
total data pairs ($i=795$ for the $H(z)$ data and $i=668$ for the
simulated GW data). Note that the sample size of the data pairs is
smaller than the original quasar sample, suffering from the limited
redshift range of the $H(z)$ measurements and the limited sample
size of the simulated GW events. However, such procedure allows
avoiding possible systematic errors brought by redshift difference.
Regarding the parametrization of the cosmic opacity, two particular
parameterizations of phenomenological $\tau(z)$ have been
extensively discussed in the above quoted papers \citep{Li13,Liao13}
\begin{eqnarray}
P1. \  \tau(z)& = & 2\epsilon z, \nonumber\\ P2. \   \tau(z)&= &
(1+z)^{2\epsilon}-1.
\end{eqnarray}
where $\epsilon$ is a constant to be constrained by observational
data. One should expect $\epsilon=0$ to be the best-fit parameters
in the confidence contours, if it is consistent with photon
conservation and there is no visible violation of the transparency
of the universe.

\begin{table*}
\caption{\label{tab:result} Summary of the constraints on the
cosmic-opacity parameter $\epsilon$ and the quasar parameters
($\beta$, $\gamma$, $\delta$), in the framework of two $\tau(z)$
parameterizations with two Hubble constant priors (see the text for
the details).}
\begin{center}
\begin{tabular}{l|l|l|l|llllll}\hline\hline
Parametrization [Data] ($H_0$ prior)   & \,\,\,\,\,\,\,\,\,\,\,\,\,\,\,\,  $\epsilon$ & \,\,\,\,\,\,\,\,\,\,\,\,\,\,\,\,$\beta$ & \,\,\,\,\,\,\,\,\,\,\,\,\,\,\,\, $\gamma$ & \,\,\,\,\,\,\,\,\,\,\,\,\,\,\,\, $\delta$    \\
\hline P1 [QSO+H(z)] (Planck $H_0$)  & $\epsilon=0.102^{+0.069}_{-0.076}$ & $\beta=7.738^{+0.542}_{-0.595}$ & $\gamma=0.616^{+0.021}_{-0.018}$& $\delta=0.225^{+0.006}_{-0.006}$    \\
  \hline
P2 [QSO+H(z)] (Planck $H_0$) & $\epsilon=0.083^{+0.111}_{-0.175}$ & $\beta=7.430^{+0.656}_{-0.623}$ & $\gamma=0.627^{+0.022}_{-0.023}$& $\delta=0.225^{+0.006}_{-0.006}$     \\
  \hline
P1 [QSO+H(z)] (Local $H_0$) & $\epsilon=0.086^{+0.072}_{-0.075}$ & $\beta=7.704^{+0.612}_{-0.620}$ & $\gamma=0.617^{+0.021}_{-0.021}$& $\delta=0.225^{+0.006}_{-0.006}$    \\
  \hline
P2 [QSO+H(z)] (Local $H_0$)  & $\epsilon=0.123^{+0.098}_{-0.153}$ & $\beta=7.484^{+0.645}_{-0.625}$ & $\gamma=0.624^{+0.022}_{-0.022}$& $\delta=0.225^{+0.006}_{-0.006}$     \\
\hline

P1 [QSO+GW] (Planck $H_0$)  & $\epsilon=0.108^{+0.045}_{-0.049}$&  $\beta=7.716^{+0.586}_{-0.572}$& $\gamma=0.617^{+0.019}_{-0.020}$& $\delta=0.225^{+0.007}_{-0.007}$   \\
  \hline
P2 [QSO+GW] (Planck $H_0$)  & $\epsilon=0.120^{+0.066}_{-0.090}$&  $\beta=7.561^{+0.604}_{-0.618}$& $\gamma=0.622^{+0.021}_{-0.020}$& $\delta=0.225^{+0.007}_{-0.006}$      \\

\hline\hline
\end{tabular}
\end{center}
\end{table*}

Let's start with the GP reconstructed $D_L(z)$ from the Hubble
parameter measurements. For the first $\tau(z)$ parametrization P1,
we obtain the best-fitted value for the cosmic-opacity parameter:
$\epsilon=0.102^{+0.069}_{-0.076}$ and
$\epsilon=0.086^{+0.072}_{-0.075}$ (at 68.3\% confidence level) with
two different priors on $H_0$ from the latest Planck CMB
observations and local Hubble constant measurements, respectively.
Working on the second $\tau(z)$ parametrization P2, the best fit
takes place at $\epsilon=0.083^{+0.111}_{-0.175}$ and
$\epsilon=0.123^{+0.098}_{-0.153}$ (at 68.3\% confidence level). In
general, even though our analysis supports a slight preference for
the non-zero cosmic opacity at $z\sim 2$, the results are still
consistent with zero cosmic opacity within $1\sigma$ confidence
level. Graphic representations and numerical results of the
constraints on the opacity parameter $\epsilon$ and quasar nuisance
parameters ($\beta$, $\gamma$, $\delta$) for P1 and P2
parameterizations are shown in Fig.~4-5 and Table 1. The significant
effect of the nonlinear relation between X-ray and UV emissions of
quasars should be pointed out: our results reveal a strong
degeneracy between the cosmic opacity parameter ($\epsilon$) and the
quasar parameters ($\beta$ and $\gamma$), i.e., a larger slope
parameter $\gamma$ and a lower intercept parameter $\beta$ will lead
to a transparent unverse ($\epsilon=0$). Such tendency has been
extensively discussed in the framework of the SNe Ia data, which
confirmed the correlation between the cosmic opacity parameter and
the absolute B-band magnitude of SNe Ia \citep{Qi19b}.

\begin{figure}
\centering
\includegraphics[scale=0.5]{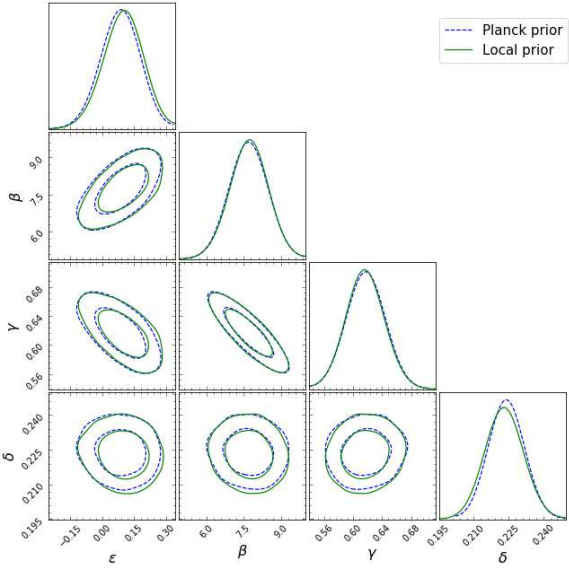}
\caption{Parametrization P1 -- constraints on the cosmic opacity
parameter $\epsilon$ and nuisance parameters ($\beta$, $\gamma$,
$\delta$), using the quasar sample and GP reconstructed $D_L(z)$
from the $H(z)$ measurements. Blue dashed line denotes the prior on
$H_0$ from Planck CMB data, while green solid line represents the
$H_0$ prior from local Hubble constant measurement. }\label{fig1}
\end{figure}

\begin{figure}
\centering
\includegraphics[scale=0.5]{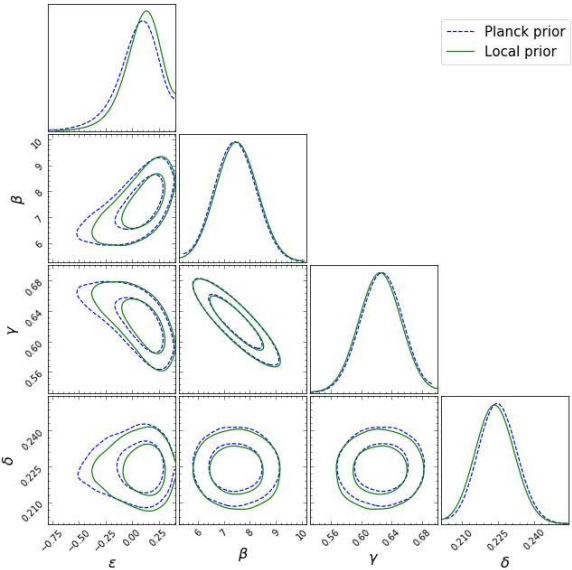}
\caption{The same as Fig.~4, but for the P2 parametrization. }
\label{fig2}
\end{figure}

\begin{figure}
\centering
\includegraphics[scale=0.35]{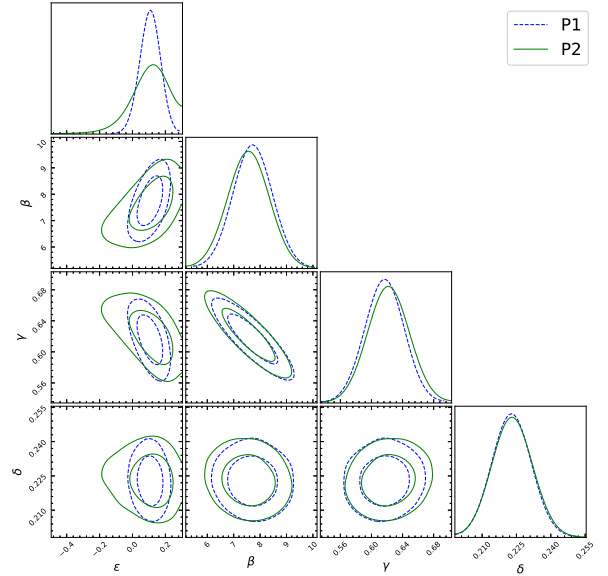}
\caption{Constraints on the opacity parameter $\epsilon$ and quasar
nuisance parameters ($\beta$, $\gamma$, $\delta$) for the P1 and P2
parameterizations, using observations of quasars and simulated GW
events. }
\end{figure}

Currently, the Hubble parameter measurements $H(z)$ are available
only up to the redshift of $z\sim 2$, therefore we perform similar
tests of cosmic opacity by combining the quasar sample and the
simulated GW data achievable with future detectors. The advantage of
such data combination lies in the fact that more quasars can be used
to investigate the opacity of the early universe ($z\sim 5$), in the
framework of the cosmological-model-independent method proposed in
this paper. Focusing on the cosmic opacity parameter $\epsilon$ and
quasar parameters ($\beta$, $\gamma$, $\delta$), we give the
1$\sigma$ and 2$\sigma$ contours for the joint distributions of any
two parameters in Fig.~6, with the corresponding numerical results
displayed in Table 1. For the first $\tau(z)$ parametrization, we
get the marginalized 1$\sigma$ constraints of the parameters:
$\epsilon=0.108^{+0.045}_{-0.049}$, $\beta=7.716^{+0.586}_{-0.572}$,
$\gamma=0.617^{+0.019}_{-0.020}$, and
$\delta=0.225^{+0.007}_{-0.007}$. In the case of second $\tau(z)$
parametrization, the final derived model parameters are:
$\epsilon=0.120^{+0.066}_{-0.090}$, $\beta=7.561^{+0.604}_{-0.618}$,
$\gamma=0.622^{+0.021}_{-0.020}$, $\delta=0.225^{+0.007}_{-0.006}$.
Compared with the previous results obtained in a low redshift range
($z\sim2$), our analysis results are still consistent with an almost
transparent universe at much higher redshifts ($z\sim 5$).
Therefore, there is no significant deviation from zero cosmic
opacity at the current observational data level, which is supported
by the simulated future GW data set and quasars flux measurements
within 2$\sigma$ confidence level. Benefit from the increase of the
number of QSO/GW pairs satisfying the redshift selection criteria, a
considerable amount of high-redshift quasars are included into the
final sample. Actually, such a combination of quasars and GWs will
enable us to get more precise measurements at the level of $\Delta
\epsilon =0.05$ and $\Delta\epsilon=0.08$ for P1 and P2
parameterizations. Meanwhile, one should also note that the derived
value of the cosmic opacity $\epsilon$ is noticeably sensitive to
the parametrization of $\tau(z)$. Such conclusion, which is
different from the findings of some previous works
\citep{Li13,Liao13,Qi19b,Ma19b}, highlights the importance of
choosing a reliable parametrization to describe the optical depth
$\tau(z)$ in the early universe.

One observation should be made is that, while the central values of
$\beta$ and $\gamma$ may slightly shift with the parametrization of
$\tau(z)$, the 68\% confidence ranges are well overlapped for the
QSO+$H(z)$ and QSO+GW data combinations. As was mentioned above,
such new type of high-redshift standard candle requires proper
calibration (in a cosmological model-independent way), considering
the strong degeneracy between the opacity parameter and the quasar
parameters. One of the most recent studies
\citep{Risaliti2018} tried to derive the slope $\gamma$ with quasar
sub-samples in different redshift bins and cross-calibrate the
intercept $\beta$ with the Union2.1 SNe Ia sample. However, it
should be emphasized that the cosmic absorption not only affects the
luminosity distances derived from nonlinear relation between X-ray
and UV emissions of quasars, but also generates influences on the
measured flux of SNe Ia \citep{Li13}. Specially, considering the
fact that the flux received by the observer will be reduced by a
factor of $e^{-\tau(z)}$, the cosmic opacity enters through the
revised $L_{UV}-L_X$ relation in terms of the fluxes $F$ and two
constants ($\gamma$, $\beta$) [Eq.~(3)]. Therefore, the relation
between the UV and X-ray luminosities of quasars could be more
accurately calibrated by external low-redshift standard candles like
SNe Ia (with redshifts overlapping with quasars), if the cosmic
opacity is precisely measured in some other way (i.e., based on the
time-delay measurements of galactic-scale strong gravitational
lensing systems \citep{Ma19b}).

Finally, it is interesting to compare our results with the previous
opacity constraints involving distances from different observations.
On the one hand, many cosmological model-independent methods (such
as nearby SNe Ia method, interpolation method and smoothing method)
have been applied to reconstruct the opacity-free luminosity
distances and match the SNe Ia and $H(z)$ data at the same redshift,
with the final results that an almost transparent universe is
favored at $z\sim 1.4$ \citep{Liao13,Liao15}. Their results implied
$\epsilon=-0.01\pm0.10$ for the P1 parametrization and
$\epsilon=-0.01\pm0.12$ for the P2 parametrization, respectively
\citep{Liao13}. On the other hand, the analysis performed by
\citet{Li13} furthermore examined the cosmic opacity in a
particularly low redshift range ($z\sim0.89$), which revealed more
stringent constraints ($\epsilon=0.009\pm0.057$ for P1
parametrization and $\epsilon=0.014\pm0.070$ for P2 parametrization)
with the combination of ADDs measured in two galaxy cluster samples
and LDs from the Union2.1 SNe Ia sample. One positive conclusion one
may draw from this comparison is, the combination of quasars and
other EM/GW data could effectively extend the cosmic-opacity
measurements to much higher redshifts ($z\sim5$).

\section{Conclusions and discussion}

In this paper, we have discussed a new model-independent
cosmological test for the cosmic opacity, the presence of which
could potentially lead to a significant deviation from photon number
conservation at higher redshifts. Such phenomena is not only related
to the foundations of observational cosmology (a transparent
universe), but also dependent on some exotic mechanisms which turn
photons into unobserved particles through extragalactic magnetic
fields \citep{Bassett04a,Bassett04b,Corasaniti06}.

We consider the opacity-dependent LDs from the non-linear relation
between X-ray and UV flux measurements in the currently largest
quasar sample \citep{Risaliti2018}, together with two types of
opacity-free LDs respectively derived from the cosmic chronometer
data ($z\sim 2$) and simulated GW events based on the future GW
observations of the ET ($z\sim5$). Focusing on the cosmic opacity
parameterized in two different forms, our results show that the
cosmic-opacity parameter $\epsilon$ can be constrained with an
accuracy of $\Delta \epsilon\sim 10^{-2}$. Interestingly, the
combination of the quasar data and the $H(z)$/GW observations in
similar redshift ranges provides a novel way to confirm a
transparent universe ($\epsilon=0$ within $2\sigma$ confidence
level) at higher redshifts ($z\sim 5$). Meanwhile, the derived value
of the cosmic-opacity parameter is noticeably sensitive to the
parametrization of $\tau(z)$. Such conclusion, which is different
from the findings of some previous works
\citep{Li13,Liao13,Qi19b,Ma19b}, highlights the importance of
choosing a reliable parametrization to describe the optical depth
$\tau(z)$ in the early universe. Finally, our findings strongly
imply a degeneracy between the cosmic-opacity parameter ($\epsilon$)
and the parameters characterizing the $L_X$-$L_{UV}$ relation of
quasars (the slope $\gamma$, the intercept $\beta$).
Therefore, more vigorous and convincing constraints on the
nonlinear relation between X-ray and UV emissions of quasars may be
expected, with a combined Hubble diagram of SNe Ia \citep{Scolnic18}
and gamma-ray bursts (GRBs) \citep{Demianski17}. Such compilation of
high-redshift distance indicators, which provides the measurement of
opacity-dependent luminosity distance covering the redshift range of
$z>2$, can also be used to check the possible tension between the
flat $\Lambda$CDM model and high-redshift Hubble diagram of
supernovae, quasars, and GRBs \citep{Lusso19}.

In addition, it is necessary to analyze several sources of
systematics that may potentially bias our quantitative analysis of
the cosmic opacity. One general concern is given by the fact that
corrections of the observed UV fluxes for dust extinction is very
necessary and meaningful for the analysis of non-linear relations
between the UV and X-ray emission of quasars. Although this problem
has been recognized long time ago, the most straightforward solution
to this issue is proposed in \citet{Risaliti2015,Risaliti2018},
which indicated that the extinction or reddening on UV flux will
underestimate the UV to X-ray ratio and generate a large luminosity
distance. One should stress that the nonconservation of the photon
number due to cosmic opacity is frequency independent in the
observed frequency range. Focusing on the effect of cosmic-opacity
parameter $\epsilon$ in this paper, it mimics three physical effects
along the light path: (1) scattering from dust or free electron; (2)
astrophysical mechanisms such as gravitational lensing; (3) exotic
physics such as photon decay and photon mixing with dilaton or axion
\citep{Bassett04a,Bassett04b,Corasaniti06}. Our results demonstrate
that the cosmic opacity plays a very similar role in the
observations of both UV and X-ray fluxes, i.e., the absorption in
various wavelength bands will lead to a large value of the
normalization constant ($\beta$) and luminosity distance ($D_L$).
The positive correlation between cosmic-opacity parameter,
normalization constant, and luminosity distance can be clearly seen
from Eq.~(3) and Fig.~4-6.

As a final remark, although the constraint on cosmic opacity by
using quasars X-ray and UV flux measurements does not significantly
improve constraints obtained by using SNe Ia, yet it helps us to
gain deeper understanding of cosmic opacity in the early universe.
With a bigger, well calibrated quasar sample acting as a new type of
standard candle in the future, it is reasonable to expect that the
observational quasar data will play an increasingly important role
in the cosmic-opacity measurements at higher redshifts.

\section*{Acknowledgments}

This paper is dedicated to the 60th anniversary of the Department of
Astronomy, Beijing Normal University. This work was supported by
National Key R\&D Program of China No. 2017YFA0402600; the National
Natural Science Foundation of China under Grants Nos. 11690023, and
11633001; Beijing Talents Fund of Organization Department of Beijing
Municipal Committee of the CPC; the Strategic Priority Research
Program of the Chinese Academy of Sciences, Grant No. XDB23000000;
the Interdiscipline Research Funds of Beijing Normal University; and
the Opening Project of Key Laboratory of Computational Astrophysics,
National Astronomical Observatories, Chinese Academy of Sciences.
This work was performed in part at Aspen Center for Physics, which
is supported by National Science Foundation grant PHY-1607611. This
work was partially supported by a grant from the Simons Foundation.
M.B. is grateful for this support. He is also grateful for support
from Polish Ministry of Science and Higher Education through the
grant DIR/WK/2018/12. M.B. obtained approval of foreign talent
introducing project in China and gained special fund support of
foreign knowledge introducing project. He also gratefully
acknowledges hospitality of Beijing Normal University.


\end{document}